# Optical amplitude and phase modulation dynamics at the single-photon level in a quantum dot ridge waveguide

Galan Moody, Corey McDonald, Ari Feldman, Todd Harvey, Richard P. Mirin, and Kevin L. Silverman*

National Institute of Standards and Technology, 325 Broadway, Boulder, CO 80305, USA

*email: kevin.silverman@nist.gov

**The amplitude and phase of a material's nonlinear optical response provide insight into the underlying electronic dynamics that determine its optical properties. Phase-sensitive nonlinear spectroscopy techniques are widely implemented to explore these dynamics through demodulation of the complex optical signal field into its quadrature components; however, complete reconstruction of the optical response requires measuring both the amplitude and phase of each quadrature, which is often lost in standard detection methods. Here, we implement a heterodyne-detection scheme to fully reconstruct the amplitude and phase response of spectral hole-burning from InAs/GaAs charged quantum dots. We observe an ultra-narrow absorption profile and a corresponding dispersive lineshape of the phase, which reflect the nanosecond optical coherence time of the charged exciton transition. Simultaneously, the measurements are sensitive to electron spin relaxation dynamics on a millisecond timescale, as this manifests as a magnetic-field dependent delay of the amplitude and phase modulation. Appreciable amplitude modulation depth and nonlinear phase shift up to ~0.09×π radians (16°) are demonstrated, providing new possibilities for quadrature modulation at faint photon levels with several independent control parameters, including photon number, modulation frequency, detuning, and externally applied fields.**

## INTRODUCTION

Semiconductor quantum dots (QDs) are an excellent solid-state platform for opto-electronic, photonic, and quantum information processing devices due to their large oscillator strength and discrete density of states [1]. These attributes manifest as a highly nonlinear optical response that can be leveraged for all-optical switching and modulation with extremely low operating energy [2–8]. Within this material class, self-assembled QDs are well-suited for low-power optical communications due to their strong amplitude and phase response spanning nanoseconds (exciton recombination), milliseconds (electron spins), and seconds (nuclear spins) [9].

Characterizing a QD's optical response function across these timescales is essential for linking the intrinsic charge carrier and exciton dynamics to device performance. Nonlinear optical spectroscopy techniques are particularly effective for accessing these dynamics in single QDs and ensembles; the amplitude and phase of the nonlinear signal reveal details of optical dephasing mechanisms, recombination channels, and multi-particle interactions and correlations [10–16]. The majority of nonlinear spectroscopy techniques explore a material's optical properties by driving inter-band transitions with a control field and measuring how that control field is converted to a signal field. In turn, this dependence provides a measurement of the material's transfer function, which contains all of the information of the optical interaction. Looking at this in reverse, if the optical properties can be controlled through material design or the application of external fields, one can engineer the transfer function to produce an ideal optical modulator. Indeed, strong synergy exists between precision spectroscopy and the development of novel optical modulators, where new materials and techniques in the former drive innovation in the latter.

To fully demodulate the amplitude and phase of a QD's nonlinear optical response, careful attention to the details of the detection method is required. In typical experiments, a 'strong' pump field creates an excitation of a dipole-allowed excitonic transition, which is then sensed by a 'weak' probe field (Fig. 1(a)). The pump field is typically modulated and the probe signal is detected at the modulation frequency using a lock-in amplifier. This measurement provides the time average of the in-phase ($X$) and quadrature ($Y$) components of the signal while suppressing noise from scatter of the pump field [14,17]; however, this approach is only sensitive to the modulated transmission of the probe beam and therefore does not provide any phase information. This type of detection is analogous to standard incoherent optical communications. An enhanced technique that is widely employed is to heterodyne the transmitted probe field with a strong local oscillator (LO) and detect the radio-frequency beatnote between the probe and LO. Parallels can be drawn here with more advanced coherent optical communication techniques, such as phase, quadrature amplitude, and single-sideband modulation [18]. Advantages of heterodyne pump-probe spectroscopy include shot-noise limited detection of a weak probe beam, separation of co-linear, co-polarized pump and probe beams, and detection of the amplitude and phase of the transmitted probe. Despite these advantages, measurements of the magnitude of the *X* and *Y* quadratures do not provide any details of their dynamics, which prevents full demodulation of the nonlinear signal unless initial assumptions about the optical response are made. For example, if the system under study exhibits a non-instantaneous response to the pump field, *e.g.* a non-zero delay $\vartheta$ in Fig. 1(a), cross-talk between the *X* and *Y* components of the lock-in signal prevents complete separation of the amplitude and phase.

In this work, we demonstrate that these limitations can be circumvented by recording the waveform of both quadrature components of a continuous-wave two-color pump-probe signal from InAs/GaAs QDs—enabling complete reconstruction of amplitude and phase modulation of the nonlinear signal. The QDs are embedded in a single-mode ridge waveguide to enhance the nonlinear light-matter interaction, providing amplitude and phase sensitivity down to the single-photon level. Reconstruction of the optical response for various pump photon numbers and detunings demonstrates that appreciable and controllable amplitude modulation and nonlinear phase shift up to ~0.09×π rad. (16°) are achieved at few-pW average power levels (~1 aJ energy during the QD lifetime) without requiring an optical cavity that relies on stringent resonance-matching conditions. With typical pump modulation frequencies in the kilohertz range, the dynamics of nuclear spins residing in the QD are too slow to follow the light intensity and therefore do not directly contribute to the signal. Electron spins, on the other hand, have typical response times similar to the modulation period and are sensitive to environmental variables such as electromagnetic fields and laser detuning. Interestingly, in the presence of a moderate external magnetic field in Faraday geometry, electron spins efficiently transfer power between the two modulation quadratures, resulting in a delayed optical response that gives rise to a frequency-dependent sideband asymmetry in the radio-frequency power spectrum of the probe. Demodulation of the heterodyned signal in a fully phase-coherent manner demonstrated here enables sensitive measurements of the QD optical response across multiple timescales. In turn, the ability to control the QD's optical amplitude and phase response may find applications as a highly configurable single element quadrature modulator, which may facilitate novel optical modulation schemes such as Hilbert transformation for faint photonic communications [19].

**EXPERIMENTAL TECHNIQUE AND DEVICE STRUCTURE**

**Phase-Sensitive Pump-Probe Spectroscopy**

To illustrate the importance of resolving the amplitude and phase of both quadrature components of the modulated optical response, let us consider a pump-probe experiment implemented using an intense pump field resonant with a QD exciton transition, which partially saturates the absorption $\alpha L$ where $L$ is the sample thickness. Modulation of the pump amplitude leads to a differential change $\Delta \alpha L$ that is imparted onto the transmitted amplitude $M(t)$ of a weak quasi-resonant probe field. Through the Kramers-Kronig relations, $\Delta \alpha L$ leads to a corresponding modulation of the refractive index $\Delta n$, which introduces a nonlinear optical phase shift $\varphi(t)$ onto the probe field (Fig. 1(a)). When mixed with a frequency-shifted LO, the signal is comprised of a beatnote at the carrier frequency given by the probe-LO offset and complex sidebands at the pump modulation frequency that contain all of the signal information.

Assuming an instantaneous optical response to the pump, it is straightforward to demodulate and measure the signal using a standard dual-phase lock-in amplifier. In this case, pure amplitude modulation of the probe appears in the $X$ quadrature of the signal from the lock-in amplifier, whereas pure phase modulation primarily appears in the $Y$ quadrature (Fig. 1(b) when setting the initial phase offset $\varphi_0 = 0$). In the case of continuous-wave excitation, the magnitude of the amplitude ($\hat{M}$) and phase ($\hat{\varphi}$) modulation can be mapped out as the pump ($f_{pump}$) and probe ($f_{probe}$) frequency detuning is varied, providing insight into the spectral response of the nonlinear optical properties (Fig. 1(c)). Measurements of $\hat{M}$ have revealed details of exciton absorption, dephasing, and relaxation dynamics through coherent population oscillations and spectral hole burning [16]. Similar experiments using pulsed optical excitation have been particularly useful for characterizing the electronic spin dynamics in a single charged QD [20] and carrier heating and differential gain dynamics in QD optical amplifiers [21].

The assumption of an instantaneous optical response relies, in general, on whether the exciton and carrier dynamics are faster than the pump modulation frequency. If these timescales are comparable, then amplitude and phase modulation can be present in both components of the lock-in signal and a simple interpretation is not possible when only measuring the power of the heterodyned signal. In fact, in the case of only pure amplitude modulation, a delayed optical response can appear as a distortion or phase shift $\vartheta_M$ of the amplitude response *M(t)* with respect to the modulated pump waveform, as illustrated in Fig. 1(a). A delay $\vartheta_M$ mixes the *X* and *Y* quadratures of the signal, resulting in an apparent phase modulation arising from pure amplitude modulation. A similar argument applies for a delay $\vartheta_\varphi$ of the phase modulation waveform. Techniques in which the nonlinear signal is demodulated to obtain only the magnitudes $\widehat{M}$ and $\hat{\varphi}$, such as optical detection using a lock-in amplifier or power measurements using a radio frequency spectrum analyzer, are not sensitive to these phase delays, limiting their ability to fully separate amplitude and phase modulation contributions. In the following subsections, we discuss how we achieve full reconstruction of *M(t)* and *φ(t)* using heterodyne-detected hole-burning spectroscopy.

**Quantum Dot Ridge Waveguide Device**

The QD waveguide device examined in this work is illustrated in the schematic diagram and scanning electron microscope image in Fig. 2(a). The 1.5 µm-wide ridge waveguide consists of a single layer of InAs/GaAs QDs between two $Al_{0.2}Ga_{0.8}As$ layers grown using molecular beam epitaxy. A single optical mode is confined to the waveguide, which is patterned using optical lithography and inductively coupled plasma etching. A 50-nm thick silicon nitride layer is deposited for surface passivation and electrical insulation. Gold electrodes separated by 2 µm are patterned on both sides of the waveguide for capacitance-voltage measurements, which indicate that on average each QD is charged with a single hole from unintentional background doping introduced during the growth [15]. The sample is held in a confocal microscope setup for continuous-wave (CW) pump-probe measurements in transmission. The pump and probe are coupled into and collected from the waveguide using 0.55 numerical aperture objective lenses each mounted on a three-axis piezoelectric nanopositioner. The lenses and sample are hermetically sealed with 15 Torr of ultra-high purity helium exchange gas in a stainless steel insert, which is fixed at 4.2 K inside a liquid helium bath magneto-cryostat. The QD photoluminescence is centered at 1045 nm with an inhomogeneous linewidth of ~50 nm due to spatial variation in the QD size and shape [Fig. 2(d)].

**Experimental Setup**

A schematic diagram of the pump-probe experimental setup is depicted in Fig. 2(b). The pump is derived from the first-order diffraction of a continuous-wave laser diode (LD1, <200 kHz linewidth) through an acousto-optical modulator (AOM), which is then coupled into the magneto-cryostat using polarization-maintaining fiber. The amplitude of the pump is modulated with a sinusoidal waveform at $f_{MOD}$ = 5 kHz by controlling the driving power to the AOM. The pump wavelength, fixed at 1050 nm, is resonant with the ground state charged exciton transition of approximately 3-5 QDs. The pump power incident on the waveguide varies from 50 pW to 1 nW, which we later show is equivalent to approximately 0.5 to 10 pump photons per QD lifetime in the waveguide. The output from a second laser diode (LD2, <200 kHz linewidth) is split into two beams: a weak probe (2 pW average power) that is fiber coupled into the cryostat, and a local oscillator that is routed around the sample and recombined with the probe for heterodyne detection. The probe frequency is shifted relative to the local oscillator by $\delta f_{pr}$ = 61 MHz using an additional AOM, enabling shot-noise limited amplitude and phase measurements of the probe/local oscillator beatnote using a commercially available radio-frequency spectrum analyzer. The nonlinear differential transmission signal appears as sidebands on the 61 MHz beatnote at $f_{MOD} = \pm 5$ kHz in the power spectrum illustrated in the top panel of Fig. 2(c). The power in the sidebands can be measured as the pump-probe detuning is varied, providing information on the charged exciton dephasing and lifetime dynamics [22].

Alternatively, complete demodulation of the signal using the spectrum analyzer is also possible through real-time measurements of each quadrature, which are depicted in Fig. 1(b). These components can be expressed as $X(t) = M(t) \times \cos[\varphi(t)]$ and $Y(t) = M(t) \times \sin[\varphi(t)]$, respectively, providing complete separation and reconstruction of the amplitude and phase dynamics. Fluctuations in the relative phase between the probe and local oscillator are passively stabilized through vibration isolation and an airtight enclosure for the experiment, and phase drift is eliminated by triggering data acquisition on the rising edge of the modulated waveform driving the pump AOM.

**RESULTS AND DISCUSSION**

In principle, the upper and lower sidebands in the power spectrum arise from both amplitude and phase modulation; however, measurements of only the sideband power cannot provide the requisite phase information to uniquely distinguish these contributions, similar to optical detection using a lock-in amplifier discussed previously. To demonstrate this, we first present in Fig. 3(a) the modulation depth of the upper ($P_{USB}$, +5 kHz) and lower ($P_{LSB}$, -5 kHz) sidebands as a function of pump-probe detuning. The modulation depth is determined by comparing the power in the sidebands and the signal-LO beatnote. Maximum modulation occurs at zero detuning, and the half-width at half-maximum of a Lorentzian fit function (solid lines) yields the transform-limited charged exciton homogeneous linewidth that has been previously characterized [22]. The upper and lower sideband lineshapes are symmetric, *i.e.* their difference $\Delta_{PSB} = P_{USB}-P_{LSB}$ is zero and independent of detuning [dashed red curve in Fig. 3(a)]. Interestingly, when applying a **B** = 1.5 T magnetic field in the Faraday configuration, the sideband modulation depth increases by nearly a factor of two and the lineshapes are asymmetric, leading to a dispersive profile of $\Delta_{PSB}$ [Fig. 3(b)]; however, since both sidebands can have amplitude and phase modulation components, in principle, the origin of the asymmetry cannot be determined from measurements of the power spectrum alone owing to the lack of phase information.

This ambiguity can be circumvented through quadrature demodulation discussed in the previous section, which provides real-time measurements of the amplitude and phase shown near zero pump-probe detuning in Fig. 3(c) and Fig. 3(e), respectively. For **B** = 0 T, the waveforms (red symbols) are delayed with respect to the pump (dashed black curve) due to lag between driving the pump AOM and subsequent amplitude modulation imparted onto the pump beam. It is apparent from sinusoidal fits to *M(t)* and *φ(t)* (red solid lines) that they oscillate in phase, *i.e.* their relative delay is zero within estimated uncertainty ($\vartheta_M - \vartheta_\varphi = 0.02 \pm 0.05$ radians). We find that $\vartheta_M$ and $\vartheta_\varphi$ are independent of pump-probe detuning (red symbols in Fig. 3(d) and Fig. 3(f)). Conversely, the magnitude of the amplitude and phase (modulation depths $\hat{M}$ and $\hat{\varphi}$) exhibit absorptive and dispersive lineshapes consistent with the Kramers-Kronig relations for absorption and refractive index, shown in Fig. 4(a) (blue and red symbols, respectively). Not surprisingly, increasing the number of pump photons by nearly an order of magnitude enhances $\hat{M}$ and $\hat{\varphi}$ as shown in Fig. 4(b). The maximum phase modulation, defined as $\Delta\hat{\varphi} = \hat{\varphi}_{max}-\hat{\varphi}_{min}$, is shown in Fig. 4(d) for various pump photons in the waveguide. The data are well-described by a saturation model for a two-level system and are fit with $\Delta\hat{\varphi}(N) = \Delta\hat{\varphi}_{max}/(1 + N^{-1})$, where *N* is the photon number normalized to the QD recombination lifetime [23]. The fit provides a straightforward method for estimating the number of pump photons absorbed by each QD, since $\Delta\hat{\varphi} = \Delta\hat{\varphi}_{max}/2$ when *N* = 1. For *N* approaching 10 (~1 aJ pump energy), we observe an asymptotic value of $\Delta\hat{\varphi} = 0.29$ rad.

(0.09×π rad.), which is a factor of ~5 larger compared to single molecules and atoms [24,25] and is the same order of magnitude as the nonlinear phase shift achieved with QD photonic and micropillar cavities [2,3,6,26].

In contrast to the power measurements of the upper and lower sidebands, the advantages of full amplitude and phase demodulation are apparent when examining the change in dynamics when the external magnetic field is applied. From Fig. 3(c) and Fig. 3(e), the applied field induces a phase lead for both *M(t)* and *φ(t)* compared to the waveforms for **B** = 0 T. The phase offsets relative to the pump waveform are shown in Fig. 3(d) and Fig. 3(f) versus pump-probe detuning. Similar to the **B** = 0 T case, $\vartheta_M$ is independent of the detuning but clearly exhibits a shift $\overline{\Delta\vartheta_M} = \vartheta_M(1.5\text{ T}) - \vartheta_M(0\text{ T}) = 0.10\pm0.03$ radians averaged over all detunings. Similarly, for phase modulation we find that $\overline{\Delta\vartheta_\varphi} = \vartheta_\varphi(1.5\text{ T}) - \vartheta_\varphi(0\text{ T}) = 0.12\pm0.04$ radians. The fact that $\overline{\Delta\vartheta_M}$ and $\overline{\Delta\vartheta_\varphi}$ are equivalent within the estimated uncertainties implies that no additional modulation mechanisms are introduced by the magnetic field, i.e. the amplitude and phase are related through Kramers-Kronig. Consistent with this observation, the amplitude- and phase-modulation depths both increase by nearly a factor of two with applied field (Fig. 4(c)). We emphasize here that the dependence of the delays $\vartheta_{M,\varphi}$ on magnetic field are responsible for the sideband asymmetry shown in Fig. 3(b), which is not evident from the power measurements of the sidebands alone.

A physical picture for the magnetic field dependence of $\vartheta_{M,\varphi}$ is obtained by considering the relevant relaxation dynamics of the positively charged exciton transition under an applied magnetic field, illustrated by the four-level energy diagram in Fig. 5(a). The ground states consist of the spin-up (⇑) and spin-down (⇓) hole states with a Zeeman energy splitting $E_h = \mu_B g_h \mathbf{B}$, where $\mu_B$ is the Bohr magneton, $g_h$ is the hole *g*-factor, and **B** is the Faraday magnetic field strength. The dipole-allowed transitions connect ⇑→⇑⇓↑ and ⇓→⇑⇓↓, where the charged exciton states ⇑⇓↑ and ⇑⇓↓ are Zeeman split by $E_e = \mu_B g_e \mathbf{B}$, where $g_e$ is the electron *g*-factor. Because the Zeeman splittings $E_h$ and $E_e$ are larger than the transition homogeneous linewidth [27], for simplicity we consider the case in which our pump is resonant only with the higher energy transition ⇓→⇑⇓↓, but similar conclusions apply for QDs in which our pump is resonant with the ⇑→⇑⇓↑ transition. After optically pumping ⇓→⇑⇓↓, the charged exciton can relax through spontaneous recombination along the direct channel ⇑⇓↓→⇓ with rate $\Gamma_{sp}$ and along the indirect channel ⇑⇓↓→⇑⇓↑→⇑→⇓ mediated by electron and hole spin relaxation with rates $\gamma_e$ and $\gamma_h$.

Direct spontaneous recombination is the dominant relaxation channel because $\Gamma_{sp} \gg \gamma_e, \gamma_h$ (gigahertz versus kilohertz) [9]; however, because $f_{MOD} \approx \gamma_e$, our measurements are also sensitive to electron spin relaxation dynamics. The direct transition $\overline{\Downarrow \rightarrow \Uparrow\Downarrow\downarrow}$ leads to a differential transmission signal at the pump modulation frequency because the carriers can follow the pump dynamics for $f_{MOD} \ll \Gamma_{sp}$. The indirect transition initiated by relaxation from $\Uparrow\Downarrow\downarrow \rightarrow \Uparrow\Downarrow\uparrow$ through an electron spin flip adds an additional component to the signal with a characteristic relaxation rate $\gamma_r$ determined by the time required to repopulate the depleted $\Downarrow$ state. For $f_{MOD} \gg \gamma_r$, the electron spin in state $\Uparrow\Downarrow\downarrow$ is effectively 'frozen' and the ground state remains depleted; thus the probe is only sensitive to the direct spontaneous recombination dynamics. For $f_{MOD} \ll \gamma_r$, the system can respond fast enough to repopulate the $\Downarrow$ spin state through the indirect channel, leading to subsequent absorption of the probe along the $\Downarrow \rightarrow \Uparrow\Downarrow\downarrow$ transition. In effect, the four-level charged exciton system acts as a first-order high-pass optical filter with a cut-off frequency $f_c = \gamma_r$ determined by the indirect channel relaxation rate. A qualitative illustration of the QD filter frequency response function and phase angle imparted onto the probe is shown in Fig. 5(c) and Fig. 5(d). For $f_{MOD} > f_c$, the QD optical filter imparts a small positive phase shift $\vartheta = \tan^{-1}(f_c/f_{MOD})$. From our measurements of an average $\overline{\Delta\vartheta_{M,\varphi}} \approx 0.11$ radians discussed previously and with $f_{MOD} = 5$ kHz, we find that $f_c = 0.55$ kHz corresponding to a response time $\tau_c = 1.8$ ms. At zero magnetic field, optical spin pumping and re-pumping of the degenerate $\Uparrow \rightarrow \Uparrow\Downarrow\uparrow$ and $\Downarrow \rightarrow \Uparrow\Downarrow\downarrow$ transitions effectively eliminates the filtering behavior of the charged QD [15,28].

Insight into the primary microscopic process governing the QD filter frequency response is provided by solving the set of coupled rate equations for the charged exciton system in Fig. 5(a). Since the $\Downarrow \rightarrow \Uparrow\Downarrow\downarrow$ transition is continually pumped, the filter response is modeled by solving the rate equations taking into account only the indirect recombination pathway, resulting in the expressions below:

$$\frac{d\,|\Uparrow\Downarrow\downarrow\rangle}{dt} = -\gamma_e\,|\Uparrow\Downarrow\downarrow\rangle + \gamma_e'\,|\Uparrow\Downarrow\uparrow\rangle$$

$$\frac{d\,|\Uparrow\Downarrow\uparrow\rangle}{dt} = \gamma_e\,|\Uparrow\Downarrow\downarrow\rangle - (\gamma_e' + \Gamma_{sp})\,|\Uparrow\Downarrow\uparrow\rangle$$

$$\frac{d\,|\Uparrow\rangle}{dt} = -\gamma_h\,|\Uparrow\rangle + \gamma_h'\,|\Downarrow\rangle + \Gamma_{sp}|\Uparrow\Downarrow\uparrow\rangle$$

$$\frac{d|\Downarrow\rangle}{dt} = \gamma_h |\Uparrow\rangle - \gamma'_h |\Downarrow\rangle. \tag{1}$$

Solutions to Eq. (1) are shown in Fig. 5(b). We set the rates $\gamma'_{e/h} = \gamma_{e/h} \times \exp(-E_{e/h}/k_b T)$, where $k_b$ is the Boltzmann constant and T = 4.2 K is the sample temperature. We use electron and hole g-factors of $g_e = 0.3$ and $g_h = 4$ for the excited and ground state splitting [29], although the exact values used here have minimal impact due to the low sample temperature used in this work, i.e. the reverse spin flip rates $\gamma'_{e/h}$ do not significantly affect the system dynamics. The total indirect-path relaxation rate $\gamma_r$ is obtained from fitting the occupancy of the $\Downarrow$ ground state to a rising exponential function (solid black line in Fig. 5(b)). Previous magneto-optical spectroscopy studies have shown that hole spin relaxation is primarily due to spin-orbit-mediated phonon scattering between the Zeeman levels, resulting in a hole spin relaxation rate $\gamma_h$ that is a factor of 5-10 faster than $\gamma_e$ in a magnetic field $\mathbf{B} \gtrsim 0.5$ T [9,30]. As a result, the frequency response of the QD filter is governed primarily by the slower electron spin relaxation rate $\gamma_e$. Equation (1) is solved for $\gamma_r$ by fixing $\gamma_h = 4$ kHz taken from [33] for $\mathbf{B} = 1.5$ T and allowing $\gamma_e$ to vary as a fitting parameter. We find that $\gamma_r$ is equal to the measured QD filter cut-off frequency $f_c$ for an electron spin relaxation rate $\gamma_e = 0.57$ kHz (lifetime $\tau_e = 1.74$ ms), which supports our assertion that electron spin relaxation governs the charged exciton optical response on the kilohertz scale. The measured lifetime $\tau_e$ is in agreement with the electron spin lifetime reported in [30].

**CONCLUSION AND OUTLOOK**

We have implemented a phase-sensitive heterodyne-detection scheme for amplitude and phase measurements of the spectral hole-burning response of self-assembled InAs QDs. This technique enables complete reconstruction of the modulation dynamics of the nonlinear optical signal, providing shot-noise limited detection sensitivity across multiple timescales from nanoseconds to milliseconds. We observe an ultra-narrow spectral hole in the QD absorption profile and a corresponding dispersive lineshape of the phase response, which reflect the nanosecond optical coherence time of charged excitons in the QDs. Simultaneously, we measure electron spin dynamics on a millisecond timescale, as this manifests as a magnetic-field dependent phase offset of the amplitude and phase modulation waveforms. We emphasize that while these effects appear as frequency-dependent modulation of the sideband power in the radio-frequency spectrum of the signal, power measurements cannot provide the requisite phase information to distinguish between modulation of the amplitude and phase—an inherent advantage of time-resolved heterodyne detection.

The phase-sensitive measurements discussed here also present new opportunities for quadrature modulation at faint photon levels. We have shown that the modulated amplitude and phase dynamics can be precisely controlled by tuning several independent parameters, including the pump power, optical frequency detuning, pump modulation frequency, and through the application of an external magnetic field to manipulate the electronic wavefunctions and energy levels. In principle, the ridge waveguide design enables broadband control of the phase and amplitude limited only by the QD inhomogeneous linewidth, since the pump sets the resonance frequency. Phase shifts approaching π might be achieved by increasing the density and number of QD layers in the waveguide, but at the expense of a reduction in the efficiency [5]. Larger phase shifts at lower optical power may be possible by optimizing the optical mode overlap with the QDs and by tailoring the electronic wavefunctions using external magnetic and electric fields [22,31].

(2007).

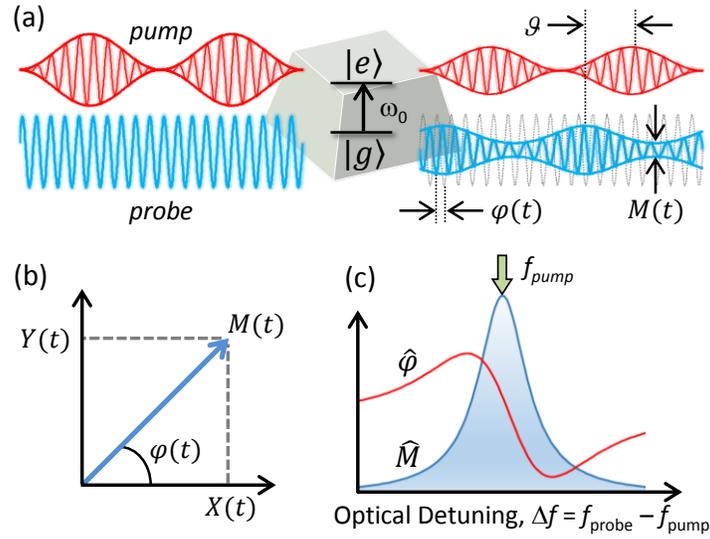

**Figure 1: (a)** Resonant optical excitation of a quantum dot charged exciton transition by a strong pump field. The modulated optical response is transferred to the amplitude $M(t)$ and phase $\varphi(t)$ of a weak probe field that can be delayed by $\vartheta$ from the modulated pump. **(b)** The amplitude and phase of the probe signal can be constructed from measurements of its in-phase ($X$) and quadrature ($Y$) components. **(c)** The magnitude of the amplitude ($\hat{M}$) and phase ($\hat{\varphi}$) modulation are related through Kramers-Kronig and are sensitive to the detuning of continuous-wave pump and probe optical frequencies, $\Delta f = f_{\text{probe}} - f_{\text{pump}}$.

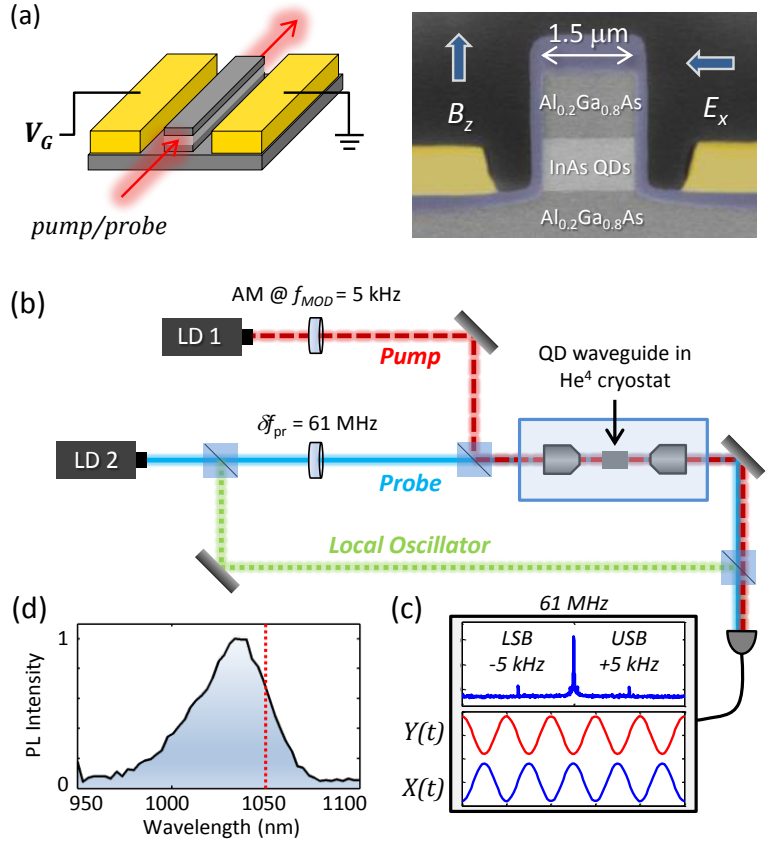

**Figure 2: (a)** Illustration (left) and false-color scanning electron microscope image (right) of the quantum dot ridge waveguide device. **(b)** Schematic diagram of the pump-probe experimental setup. The pump and probe derived from two independently tunable continuous-wave laser diodes (LD1 and LD2) are fiber coupled into the QD ridge waveguide held at 4.2 K. The beatnote at the probe/local oscillator difference frequency $\delta f_{pr}$ = 61 MHz is detected using a fast photodiode and radio-frequency spectrum analyzer shown in **(c)**. The pump amplitude is modulated at $f_{MOD}$ = 5 kHz, leading to a differential change in the amplitude and phase of the probe that appears as upper (USB) and lower (LSB) sidebands at +5 kHz and -5 kHz relative to the 61 MHz beatnote. Measurements of the in-phase ($X$) and quadrature ($Y$) components of the probe/local oscillator heterodyne signal enable demodulation of the amplitude ($M$) and phase ($\varphi$) dynamics. **(d)** An ensemble photoluminescence spectrum taken at 4.2 K. The excitation wavelength for the pump-probe experiments is indicated by the dashed line at 1050 nm.

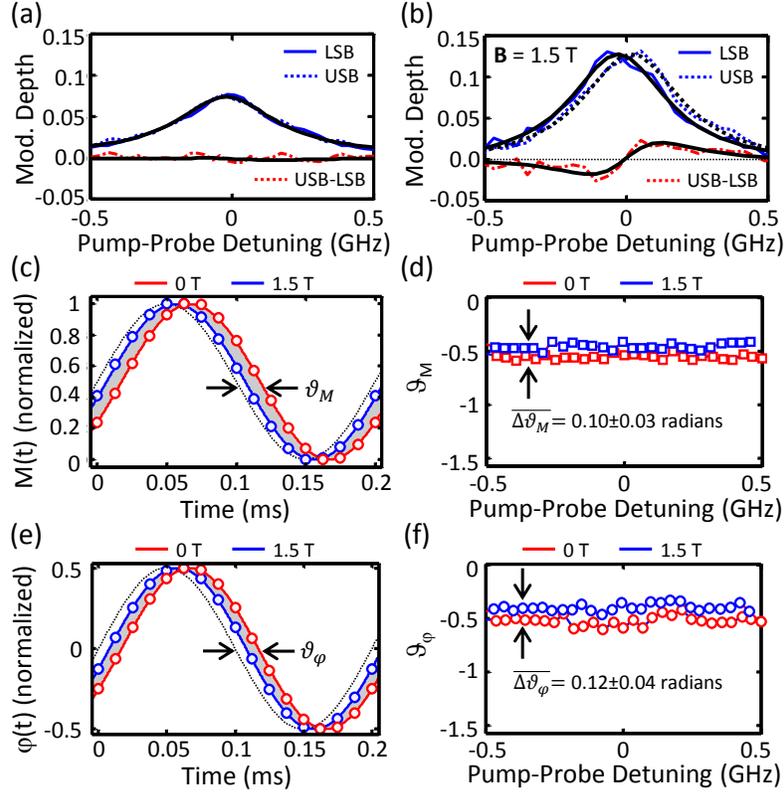

**Figure 3: (a)** Modulation depth of the lower and upper sidebands (solid and dashed blue curves, respectively) and their difference (dashed red curve). The sideband lineshapes are fit with a Lorentzian function (solid black curve). **(b)** Application of a **B** = 1.5 T external magnetic field in Faraday geometry results in frequency-dependent sideband asymmetry, which is illustrated by the dispersive lineshape of the difference in the sidebands. **(c)** The signal amplitude modulation dynamics with (blue symbols and curve) and without (red symbols and curve) an applied magnetic field. The pump amplitude modulation waveform is depicted by the dashed curve. **(d)** The delay $\vartheta_M$ between the probe amplitude modulation and the waveform driving the pump AOM. The average difference in the delay with and without the applied magnetic field is $\overline{\Delta\vartheta_M} = 0.10\pm0.03$ radians. **(e)** The phase modulation dynamics exhibit similar behavior with an average shift with applied magnetic field of $\overline{\Delta\vartheta_\varphi} = 0.12\pm0.04$ radians, shown in **(f)**.

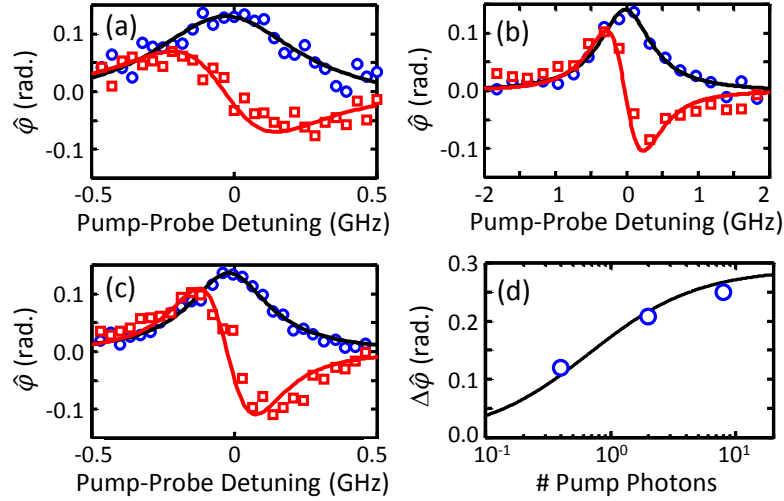

**Figure 4:** Phase modulation magnitude $\hat{\varphi}$ (red squares) is shown as a function of pump-probe detuning for **B** = 0 T and **(a)** 0.4 pump photons and **(b)** 2 pump photons and **(c)** 0.4 pump photons and **B** = 1.5 T. Normalized differential absorption lineshapes (blue circles) and Lorentzian fits are shown for reference. **(d)** Nonlinear phase shift $\Delta\hat{\varphi} = \hat{\varphi}_{max}-\hat{\varphi}_{min}$ as a function of the number of pump photons in the waveguide for **B** = 0 T. The data are fit with a saturation model for a two-level system (solid line).

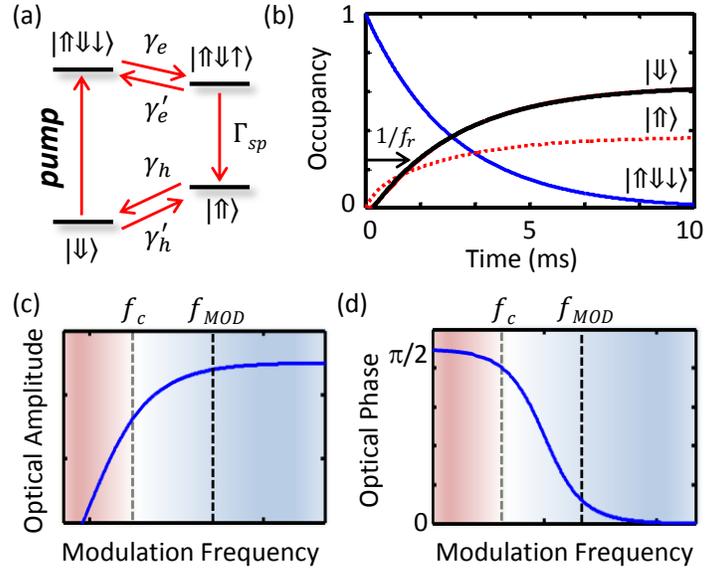

**Figure 5: (a)** Energy level diagram for the positively charged exciton with an applied magnetic field in the Faraday geometry. **(b)** Solutions to the rate equation model for the energy diagram in **(a)**. A schematic illustration of the frequency response function **(c)** and phase angle **(d)** of the output signal from the QD acting as a high-pass optical filter. The cut-off frequency $f_c$ is determined by the rate at which carriers from $\Uparrow\Downarrow\downarrow$ repopulate the ground state $\Downarrow$ through the indirect relaxation channel.